\documentclass{article}
\usepackage{spconf,amsmath,graphicx,multirow,tabularx, booktabs,array}
\usepackage{color,soul}
\usepackage{xcolor}
\usepackage{array}

\def\x{{\mathbf x}}
\def\y{{\mathbf y}}
\def\z{{\mathbf z}}

\def\Z{{\cal Z}}

\title{Transformer Transducer: A Streamable Speech Recognition Model with Transformer Encoders and RNN-T Loss\thanks{This is the final version of the paper submitted to the ICASSP 2020 on Oct 21, 2019.}}

\name{Qian Zhang, Han Lu, Hasim Sak, Anshuman Tripathi, Erik McDermott, Stephen Koo, Shankar Kumar}
\address{\{zhaqian, luha, hasim, anshumant, erikmcd, kookaburra, shankarkumar\}@google.com\\Google Inc., USA}

\begin{document}
\ninept
\maketitle
\begin{abstract}
In this paper we present an end-to-end speech recognition model with Transformer encoders that can be used in a streaming speech recognition system. Transformer computation blocks based on self-attention are used to encode both audio and label sequences independently.
The activations from both audio and label encoders are combined with a feed-forward layer to compute a probability distribution over the label space for every combination of acoustic frame position and label history. This is similar to the Recurrent Neural Network Transducer (RNN-T) model, which uses RNNs for information encoding instead of Transformer encoders. The model is trained with the RNN-T loss well-suited to streaming decoding. We present results on the LibriSpeech dataset showing that limiting the left context for self-attention in the Transformer layers makes decoding computationally tractable for streaming, with only a slight degradation in accuracy. We also show that the full attention version of our model beats the-state-of-the art accuracy on the LibriSpeech benchmarks. Our results also show that we can bridge the gap between full attention and limited attention versions of our model by attending to a limited number of future frames.
\end{abstract}
\section{Introduction}
\label{sec:intro}
In the past few years, models employing self-attention~\cite{vaswani2017attention} have achieved state-of-art results for many tasks, such as machine translation, language modeling, and language understanding~\cite{vaswani2017attention, dai2019transformer}. In particular, large Transformer-based language models have brought gains in speech recognition tasks when used for second-pass re-scoring and in first-pass shallow fusion~\cite{irie2019language}.
As typically used in sequence-to-sequence transduction tasks~\cite{transformerasr:01, transformerasr:02, transformermt:01, transformerasr:03,transformerasr:04}, Transformer-based models attend over encoder features using decoder features, implying that the decoding has to be done in a label-synchronous way, thereby posing a challenge for streaming speech recognition applications. An additional challenge for streaming speech recognition with these models is that the number of computations for self-attention increases quadratically with input sequence size. For streaming to be computationally practical, it is highly desirable that the time it takes to process each frame remains constant relative to the length of the input. Transformer-based alternatives to RNNs have recently been explored for use in ASR \cite{Povey2018ATS,Dong_2019,Sperber_2018,tsunoo2019online}.

For streaming speech recognition models, recurrent neural networks (RNNs) have been the {\em de facto} choice since they can model the temporal dependencies in the audio features effectively~\cite{Sak:14a} while maintaining a constant computational requirement for each frame. Streamable end-to-end modeling architectures such as the Recurrent Neural Network Transducer (RNN-T)~\cite{graves:12,rao2017exploring,he2017asru}, Recurrent Neural Aligner (RNA)~\cite{sak:17}, and Neural Transducer~\cite{jaitly:15} utilize an encoder-decoder based framework where both encoder and decoder are layers of RNNs that generate features from audio and labels respectively. In particular, the RNN-T and RNA models are trained to learn alignments between the acoustic encoder features and the label encoder features, and so lend themselves naturally to frame-synchronous decoding.

Several optimization techniques have been evaluated to enable running RNN-T on device~\cite{he2017asru}. In addition, extensive architecture and modeling unit exploration has been done for RNN-T~\cite{rao2017exploring}. In this paper, we explore the possibility of replacing RNN-based audio and label encoders in the conventional RNN-T architecture with Transformer encoders. With a view to preserving model streamability, we show that Transformer-based models can be trained with self-attention on a fixed number of past input frames and previous labels. This results in a degradation of performance (compared to attending to all past input frames and labels), but then the model satisfies a constant computational requirement for processing each frame, making it suitable for streaming. Given the simple architecture and parallelizable nature of self-attention computations, we observe large improvements in training time and training resource utilization compared to RNN-T models that employ RNNs.

\begin{figure}[t]
\centering
\includegraphics[width=0.45\columnwidth]{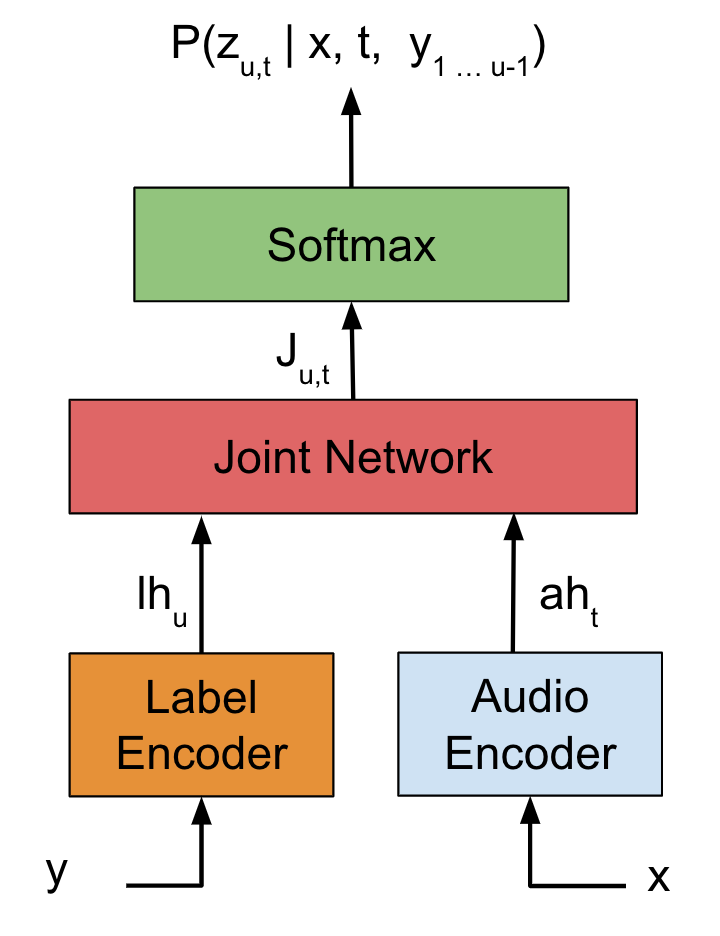}
\caption{RNN/Transformer Transducer architecture.}
\label{fig:rnnt_arch}
\end{figure}

The RNN-T architecture\footnote{We use "RNN-T architecture" or "RNN-T model" interchangeably in this paper to refer to the neural network architecture described in Eq.~(\ref{eq:enc-dec}), and Eq.~(\ref{eq:joint}), and "RNN-T loss", defined in Eq.~(\ref{eq:forward_backward_loss}), to refer to the loss used to train this architecture.} (as depicted in Figure~\ref{fig:rnnt_arch}) is a neural network architecture that can be trained end-to-end with the RNN-T loss to map input sequences (e.g. audio feature vectors) to target sequences (e.g. phonemes, graphemes).
Given an input sequence of real-valued vectors of length $T$, $\x = (x_1, x_2, ..., x_T)$, the RNN-T model tries to predict the target sequence of labels $\y = (y_1, y_2, ..., y_U)$ of length $U$.

Unlike a typical attention-based sequence-to-sequence model, which attends over the entire input for every prediction in the output sequence, the RNN-T model gives a probability distribution over the label space at every time step, and the output label space includes an additional null label to indicate the lack of output for that time step  --- similar to the Connectionist Temporal Classification (CTC) framework~\cite{graves:06}. But unlike CTC, this label distribution is also conditioned on the previous label history.

The RNN-T model defines a conditional distribution $P(\z|\x)$ over all the possible alignments, where
$$\z = [(z_1, t_1), (z_2, t_2), ..., (z_{\overline{U}}, t_{\overline{U}})]$$ is a sequence of $(z_i, t_i)$ pairs of length $\overline{U}$, and $(z_i, t_i)$ represents an alignment between output label $z_i$ and the encoded feature at time $t_i$. The labels $z_i$ can optionally be blank labels (null predictions). Removing the blank labels gives the actual output label sequence $\y$, of length $U$.

We can marginalize $P(\z|\x)$ over all possible alignments $\z$ to obtain the probability of the target label sequence $\y$ given the input sequence $\x$,
\begin{equation} \label{eq:loss}
P(\y|\x) = \sum_{\z \in \Z(\y,T)}{P(\z|\x)},
\end{equation}
where $\Z(\y,T)$ is the set of valid alignments of length $T$ for the label sequence.

\section{Transformer Transducer}
\subsection{RNN-T Architecture and Loss}
\label{sec:rnnt}

In this paper, we present all experimental results with the RNN-T loss~\cite{graves:12} for consistency, which performs similarly to the monotonic RNN-T loss~\cite{tripathi:19a} in our experiments.

The probability of an alignment $P(\z|\x)$ can be factorized as
\begin{align}
    P(\z|\x) = \prod_i P(z_i|\x, t_i, \mathrm{Labels}(z_{1:(i-1)})),
\end{align} 
where $\mathrm{Labels}(z_{1:(i-1)})$ is the sequence of non-blank labels in $z_{1:(i-1)}$. The RNN-T architecture parameterizes $P(\z|\x)$ with an audio encoder, a label encoder, and a joint network. The encoders are two neural networks that encode the input sequence and the target output sequence, respectively. Previous work~\cite{graves:12} has employed Long Short-term Memory models (LSTMs) as the encoders, giving the RNN-T its name. However, this framework is not restricted to RNNs. In this paper, we are particularly interested in replacing the LSTM encoders with Transformers~\cite{vaswani2017attention, dai2019transformer}. In the following, we refer to this new architecture as the Transformer Transducer (T-T). As in the original RNN-T model, the joint network combines the audio encoder output at $t_i$ and the label encoder output given the previous non-blank output label sequence $\mathrm{Labels}(z_{1:(i-1)})$ using a feed-forward neural network with a softmax layer, inducing a distribution over the labels. The model defines $P(z_i|\x, t_i, \mathrm{Labels}(z_{1:(i-1)}))$ as follows:
\begin{equation}
\begin{split}\label{eq:enc-dec}
\mathrm{Joint} = &\mathrm{Linear}(\mathrm{AudioEncoder}_{t_{i}}(\x)) +  \\
        &\mathrm{Linear}(\mathrm{LabelEncoder}(\mathrm{Labels}(z_{1:(i-1)}))))
\end{split}
\end{equation}
\begin{equation}
\begin{split} \label{eq:joint}
P(z_i|\x, t_i, \mathrm{Labels}(&z_{1:(i-1))}) = \\
      &\mathrm{Softmax}(\mathrm{Linear}(\mathrm{tanh}(\mathrm{Joint}))),
\end{split}
\end{equation}
where each $\mathrm{Linear}$ function is a different single-layer feed-forward neural network, $\mathrm{AudioEncoder}_{t_{i}}(\x)$ is the audio encoder output at time $t_i$, and $\mathrm{LabelEncoder}(\mathrm{Labels}(z_{1:(i-1)}))$ is the label encoder output given the previous non-blank label sequence.

To compute Eq.~(\ref{eq:loss}) by summing all valid alignments naively is computationally intractable. Therefore, we define the forward variable $\alpha(t,u)$ as the sum of probabilities for all paths ending at time-frame $t$ and label position $u$. We then use the forward algorithm \cite{graves:12,rabiner:93} to compute the last alpha variable $\alpha({T, U})$, which corresponds to $P(\y|\x)$ defined in Eq.~(\ref{eq:loss}). Efficient computation of $P(\y|\x)$ using the forward algorithm is enabled by the fact that the local probability estimate (Eq.~(\ref{eq:joint})) at any given label position and any given time-frame is not dependent on the alignment~\cite{graves:12}. The training loss for the model is then the sum of the negative log probabilities defined in Eq.~(\ref{eq:loss}) over all the training examples, 
\begin{equation} \label{eq:forward_backward_loss}
    \textrm{loss} = -\sum_i \log P(\y_i|\x_i) = - \sum_i \alpha(T_i, U_i),
\end{equation}
where $T_i$ and $U_i$ are the lengths of the input sequence and the output target label sequence of the $i$-th training example, respectively.

\subsection{Transformer}
\begin{figure}[t]

    \centering
    \includegraphics[width=9cm,height=7cm]{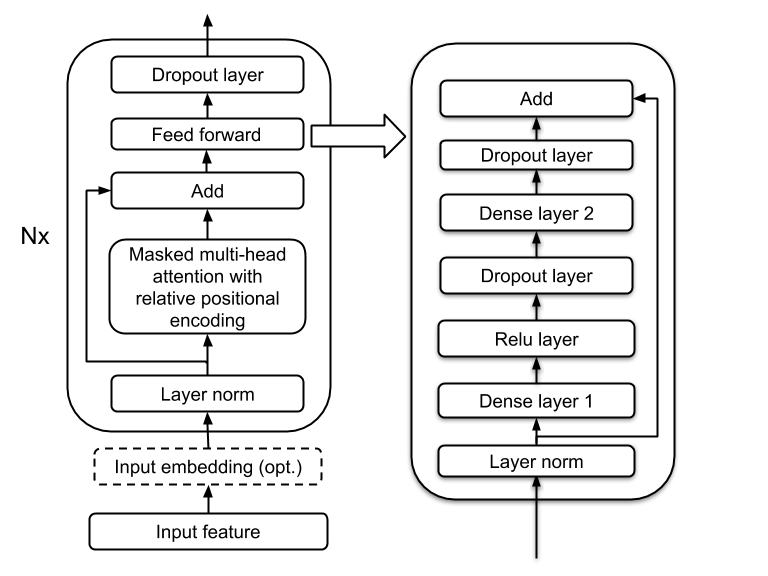}
    \caption{Transformer encoder architecture.}
    \label{fig:transformer_arch}
    \vspace{-11pt} 
\end{figure}

The Transformer~\cite{vaswani2017attention} is composed of a stack of multiple identical layers. Each layer has two sub-layers, a multi-headed attention layer and a feed-forward layer. Our multi-headed attention layer first applies $\mathrm{LayerNorm}$, then projects the input to $\mathrm{Query}$, $\mathrm{Key}$, and $\mathrm{Value}$ for all the heads~\cite{dai2019transformer}. The attention mechanism is applied separately for different attention heads. The attention mechanism provides a flexible way to control the context that the model uses. For example, we can mask the attention score to the left of the current frame to produce output conditioned only on the previous state history. The weight-averaged $\mathrm{Value}$s for all heads are concatenated and passed to a dense layer. We then employ a residual connection on the normalized input and the output of the dense layer to form the final output of the multi-headed attention sub-layer (i.e. $\mathrm{LayerNorm}(x) + \mathrm{AttentionLayer}(\mathrm{LayerNorm}(x))$, where $x$ is the input to the multi-headed attention sub-layer). We also apply dropout on the output of the dense layer to prevent overfitting. Our feed-forward sub-layer applies $\mathrm{LayerNorm}$ on the input first, then applies two dense layers. We use $\mathrm{ReLu}$ as the activation for the first dense layer. Again, dropout to both dense layers for regularization, and a residual connection of normalized input and the output of the second dense layer (i.e. $\mathrm{LayerNorm}(x) + \mathrm{FeedForwardLayer}(\mathrm{LayerNorm}(x))$, where $x$ is the input to the feed-forward sub-layer) are applied. See Figure~\ref{fig:transformer_arch} for more details. 

Note that $\mathrm{LabelEncoder}$ states do not attend to $\mathrm{AudioEncoder}$ states, in contrast to the architecture in~\cite{vaswani2017attention}. As discussed in the Introduction, doing so poses a challenge for streaming applications. Instead, we implement $\mathrm{AudioEncoder}$ and $\mathrm{LabelEncoder}$ in Eq.~(\ref{eq:enc-dec}), which are LSTMs in conventional RNN-T architectures \cite{graves:12,he2017asru,rao2017exploring}, using the Transformers described above. In tandem with the RNN-T architecture described in the previous section, the attention mechanism here only operates within $\mathrm{AudioEncoder}$ or $\mathrm{LabelEncoder}$, contrary to the standard practice for Transformer-based systems. In addition, so as to model sequential order, we use the relative positional encoding proposed in~\cite{dai2019transformer}. With relative positional encoding, the encoding only affects the attention score instead of the $\mathrm{Value}$s being summed. This allows us to reuse previously computed states rather than recomputing all previous states and getting the last state in an overlapping inference manner when the number of frames or labels that $\mathrm{AudioEncoder}$ or $\mathrm{LabelEncoder}$ processed is larger than the maximum length used during training (which would again be intractable for streaming applications). More specifically, the complexity of running one-step inference to get activations at time $t$ is $\mathrm{O}(t)$, which is the computation cost of attending to $t$ states and of the feed-forward process for the current step when using relative positional encoding. On the other hand, with absolute positional encoding, the encoding added to the input should be shifted by one when $t$ is larger than the maximum length used during training, which precludes re-use of the states, and makes the complexity $\mathrm{O}(t^2)$. However, even if we can reduce the complexity from $\mathrm{O}(t^2)$ to $\mathrm{O}(t)$ with relative positional encoding, there is still the issue of latency growing over time. One intuitive solution is to limit the model to attend to a moving window $W$ of states, making the one-step inference complexity constant. Note that training or inference with attention to limited context is not possible for Transformer-based models that have attention from $\mathrm{Decoder}$ to $\mathrm{Encoder}$, as such a setup is itself trying to learn the alignment. In contrast, the separation of $\mathrm{AudioEncoder}$ and $\mathrm{LabelEncoder}$, and the fact that the alignment is handled by a separate forward-backward process, within the RNN-T architecture, makes it possible to train with attention over an explicitly specified, limited context.

\begin{table}[t]
    \centering
    \caption{Transformer encoder parameter setup.}
    \begin{tabular}{c|c}
    \toprule
    Input feature/embedding size &  512 \\
    Dense layer 1 & 2048 \\
    Dense layer 2 & 1024 \\
    Number attention heads & 8 \\
    Head dimension & 64  \\ 
    Dropout ratio & 0.1  \\
    \bottomrule
    \end{tabular}
    \label{tab:tansformer_paramters}
\end{table}

\begin{table}[t]
    \centering
    \caption{Comparison of WERs for Hybrid (streamable), LAS (e2e), RNN-T (e2e \& streamable) and Transformer Transducer models (e2e \& streamable) on LibriSpeech test sets.}
    \begin{tabular}{c|c|c|c|c|c|}
    \toprule
    \multirow{2}*{Model} & Param &  \multicolumn{2}{c}{No LM (\%)} & \multicolumn{2}{c}{With LM (\%)} \\
    & size  & clean & other & clean & other  \\
    \midrule
     Hybrid {\cite{Wang2019hybrid}} & - & - & - & 2.26 & 4.85 \\
     LAS{\cite{park2019specaugment}} & 361M & 2.8 & 6.8 & 2.5 & 5.8 \\
     BiLSTM RNN-T    & 130M & 3.2 & 7.8  & -  & - \\ 
     FullAttn T-T (Ours)  &  139M & 2.4 & 5.6 & \textbf{2.0} & \textbf{4.6} \\
    \bottomrule
    \end{tabular}
    \label{tab:comp}
\end{table}

\section{Experiments and Results}

\subsection{Data}
We evaluated the proposed model using the publicly available LibriSpeech ASR corpus~\cite{panayotov2015librispeech}.
The LibriSpeech dataset consists of 970 hours of audio data with corresponding text transcripts (around 10M word tokens) and an additional 800M word token text only dataset. The paired audio/transcript dataset was used to train T-T models and an LSTM-based baseline. The full 810M word tokens text dataset was used for standalone language model (LM) training. We extracted 128-channel logmel energy values from a 32 ms window, stacked every 4 frames, and sub-sampled every 3 frames, to produce a 512-dimensional acoustic feature vector with a stride of 30 ms. Feature augmentation~\cite{park2019specaugment} was applied during model training to prevent overfitting and to improve generalization, with only frequency masking ($\mathrm{F}=50$, $\mathrm{mF}=2$) and time masking ($\mathrm{T}=30$, $\mathrm{mT}=10$).

\subsection{Transformer Transducer}
Our Transformer Transducer model architecture has 18 audio and 2 label encoder layers. Every layer is identical for both audio and label encoders. The details of computations in a layer are shown in Figure~\ref{fig:transformer_arch} and Table ~\ref{tab:tansformer_paramters}. All the models for experiments presented in this paper are trained on 8x8 TPU with a per-core batch size of 16 (effective batch size of 2048). The learning rate schedule is ramped up linearly from 0 to $2.5\mathrm{e}{-4}$ during first 4K steps, it is then held constant till 30K steps and then decays exponentially to $2.5\mathrm{e}{-6}$ till 200K steps. During training we also added a gaussian noise($\mu=0,\sigma=0.01$) to model weights~\cite{NIPS2011_4329} starting at 10K steps. We train this model to output grapheme units in all our experiments.
We found that the Transformer Transducer models trained much faster ($\approx 1$ day) compared to the an LSTM-based RNN-T model ($\approx 3.5$ days), with a similar number of parameters.

\begin{table}[t]
    \centering
    \caption{Limited left context per layer for audio encoder.}
    \begin{tabular}{p{0.5cm} p{0.5cm}| p{1.5cm} p{1.3cm}|p{1.3cm}}
    \toprule
     \multicolumn{2}{c}{Audio Mask} & Label Mask & \multicolumn{2}{c}{WER (\%)} \\
     left & right &  \multicolumn{1}{c|}{left} &  Test-clean & Test-other\\ 
     \hline
     10 & 0 & \multicolumn{1}{c|}{20}  & \multicolumn{1}{c|}{4.2} & \multicolumn{1}{c}{11.3} \\
     6 & 0 &  \multicolumn{1}{c|}{20}  & \multicolumn{1}{c|}{4.3} & \multicolumn{1}{c}{11.8} \\
     2 & 0 &  \multicolumn{1}{c|}{20}  & \multicolumn{1}{c|}{4.5} & \multicolumn{1}{c}{14.5} \\
    \bottomrule
    \end{tabular}
    \label{tab:context_audio_left}
\end{table}

\label{sec:results}
\subsection{Results}
We first compared the performance of Transformer Transducer (T-T) models with full attention on audio to an RNN-T model using a bidirectional LSTM audio encoder. As shown in Table~\ref{tab:comp}, the T-T model significantly outperforms the LSTM-based RNN-T baseline. We also observed that T-T models can achieve competitive recognition accuracy with existing wordpiece-based end-to-end models with similar model size. To compare with systems using shallow fusion \cite{graves:06,chorowski:17} with separately trained LMs, we also trained a Transformer-based LM with the same architecture as the label encoder used in T-T, using the full 810M word token dataset. This Transformer LM (6 layers; 57M parameters) had a perplexity of $2.49$ on the {\em dev-clean} set; the use of dropout, and of larger models, did not improve either perplexity or WER.  Shallow fusion was then performed using that LM and both the trained T-T system and the trained bidirectional LSTM-based RNN-T baseline, with scaling factors on the LM output and on the non-blank symbol sequence length tuned on the LibriSpeech dev sets. The results are shown in Table~\ref{tab:comp} in the ``With LM'' column. The shallow fusion result for the T-T system is competitive with corresponding results for top-performing existing systems. 

\begin{figure}[t]
    \includegraphics[width=8.5cm]{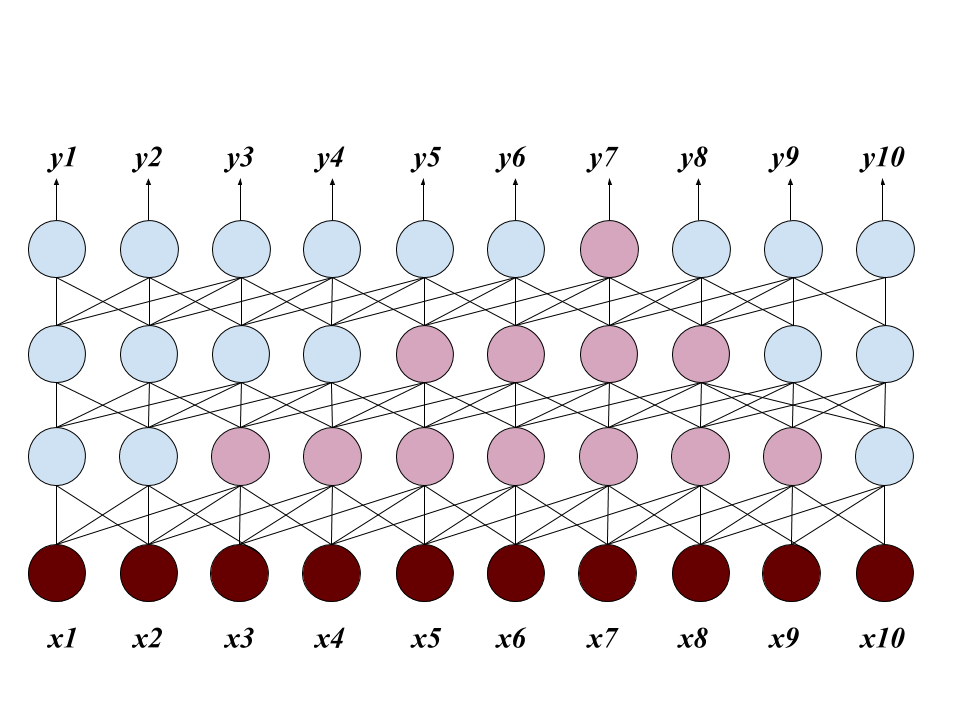}
    \caption{Transformer context masking for the $y_7$ position (left=2, right=1)}
    \label{fig:transformer_context}
\end{figure}

\begin{table}[t]
    \centering
    \caption{Limited right context per layer for audio encoder.}
    \begin{tabular}{p{0.5cm} p{0.5cm}| p{1.5cm}  p{1.3cm}|p{1.3cm}}
    \toprule
     \multicolumn{2}{c}{Audio Mask}  & Label Mask  &  \multicolumn{2}{c}{WER (\%)} \\
     left & right & \multicolumn{1}{c|}{left} & Test-clean & Test-other \\ 
     \hline
     512 & 512 & \multicolumn{1}{c|}{20} & \multicolumn{1}{c|}{2.4} & \multicolumn{1}{c}{5.6} \\
     512& 10 & \multicolumn{1}{c|}{20} & \multicolumn{1}{c|}{2.7} & \multicolumn{1}{c}{6.6} \\
     512 & 6 & \multicolumn{1}{c|}{20} & \multicolumn{1}{c|}{2.8} & \multicolumn{1}{c}{6.9} \\
     512 & 2 & \multicolumn{1}{c|}{20} & \multicolumn{1}{c|}{3.0} & \multicolumn{1}{c}{7.7} \\
     10 & 0 & \multicolumn{1}{c|}{20} & \multicolumn{1}{c|}{4.2} & \multicolumn{1}{c}{11.3} \\

    \bottomrule
    \end{tabular}
    \label{tab:context_audio_right}
\end{table}

\begin{table}[t]
    \centering
     \caption{Limited left context per layer for label encoder.}
    \begin{tabular}{p{0.5cm} p{0.5cm}| p{1.5cm}  p{1.3cm}|p{1.3cm}}
    \toprule
      \multicolumn{2}{c}{Audio Mask} & \multicolumn{1}{c}{Label Mask}  &  \multicolumn{2}{c}{WER (\%)}   \\ 
     left & right & \multicolumn{1}{c|}{left} & Test-clean & Test-other \\ 
     \hline
     10 &  0 & \multicolumn{1}{c|}{20}  & \multicolumn{1}{c|}{4.2} & \multicolumn{1}{c}{11.3} \\
     10 &  0 & \multicolumn{1}{c|}{4} & \multicolumn{1}{c|}{4.2} & \multicolumn{1}{c}{11.4} \\
     10 &  0 &  \multicolumn{1}{c|}{3} & \multicolumn{1}{c|}{4.2} & \multicolumn{1}{c}{11.4} \\
     10 &  0 &  \multicolumn{1}{c|}{2} & \multicolumn{1}{c|}{4.3} & \multicolumn{1}{c}{11.5}  \\
     10 &  0 & \multicolumn{1}{c|}{1} & \multicolumn{1}{c|}{4.4} & \multicolumn{1}{c}{12}\\

    \bottomrule
    \end{tabular}
    \label{tab:context_label}
\end{table}

\begin{table}[!htbp]
    \centering
    \caption{Results for limiting audio and label context for streaming.}
    \begin{tabular}{p{0.5cm} p{0.5cm}| p{1.5cm}  p{1.3cm}|p{1.3cm}}
    \toprule
     \multicolumn{2}{c}{Audio Mask} & Label Mask   & \multicolumn{2}{ c}{WER (\%)}  \\ 
     left & right & \multicolumn{1}{c|}{left} & Test-clean & Test-other \\ 
     \hline
     512 & 512 & \multicolumn{1}{c|}{20} & \multicolumn{1}{c|}{ 2.4} & \multicolumn{1}{c}{5.6} \\
     10 & 2 & \multicolumn{1}{c|}{2} & \multicolumn{1}{c|}{3.6} & \multicolumn{1}{c}{10} \\
     10 & 0 & \multicolumn{1}{c|}{20} & \multicolumn{1}{c|}{4.2} & \multicolumn{1}{c}{11.3} \\

    \bottomrule
    \end{tabular}
    \label{tab:context_final}
\end{table}

Next, we ran training and decoding experiments using T-T models with limited attention windows over audio and text, with a view to building online streaming speech recognition systems with low latency. Similarly to the use of unidirectional RNN audio encoders in online models, where activations for time $t$ are computed with conditioning only on audio frames before $t$, here we constrain the $\mathrm{AudioEncoder}$ to attend to the left of the current frame by masking the attention scores to the right of the current frame. In order to make one-step inference for $\mathrm{AudioEncoder}$ tractable (i.e. to have constant time complexity), we further limit the attention for $\mathrm{AudioEncoder}$ to a fixed window of previous states by again masking the attention score. Due to limited computation resources, we used the same mask for different Transformer layers, but the use of different contexts (masks) for different layers is worth exploring. The results are shown in Table~\ref{tab:context_audio_left}, where N in the first two columns indicates the number of states that the model uses to the left or right of the current frame. As we can see, using more audio history gives the lower WER, but considering a streamable model with reasonable time complexity for inference, we experimented with a left context of up to 10 frames per layer.

Similarly, we explored the use of limited right context to allow the model to see some future audio frames, in the hope of bridging the gap between a streamable T-T model (left = 10, right = 0) and a full attention T-T model (left = 512, right = 512). Since we apply the same mask for every layer, the latency introduced by using right context is aggregated over all the layers. For example, in Figure~\ref{fig:transformer_context}, to produce $y_7$ from a 3-layer Transformer with one frame of right context, it actually needs to wait for $x_{10}$ to arrive, which is 90 ms latency in our case. To explore the right context impact for modeling, we did comparisons with fixed 512 frames left context per layer to compared with full attention T-T model. As we can see from Table~\ref{tab:context_audio_right}, with right context of 6 frames per layer (around 3.2 secs of latency), the performance is around 16\% worse than full attention model. Compared with streamable T-T model, 2 frames right context per layer (around 1 sec of  latency) brings around 30\% improvements.

In addition, we evaluated how the left context used in the T-T $\mathrm{LabelEncoder}$ affects performance. In Table~\ref{tab:context_label}, we show that constraining each layer to only use three previous label states yields the similar accuracy with the model using 20 states per layer. It shows very limited left context for label encoder is good engough for T-T model. We see a similar trend when limiting left label states while using a full attention T-T audio encoder.

Finally, Table~\ref{tab:context_final} reports the results when using a limited left context of 10 frames, which reduces the time complexity for one-step inference to a constant, with look-ahead to future frames, as a way of bridging the gap between the performance of left-only attention and full attention models.

\section{Conclusions}
\label{sec:conclusions}
In this paper, we presented the Transformer Transducer model, embedding Transformer based self-attention for audio and label encoding within the RNN-T architecture, resulting in an end-to-end model that can be optimized using a loss function that efficiently marginalizes over all possible alignments and that is well-suited to time-synchronous decoding. This model achieves a new state-of-the-art accuracy on the LibriSpeech benchmark, and can easily be used for streaming speech recognition by limiting the audio and label context used in self-attention. Transformer Transducer models train significantly faster than LSTM based RNN-T models, and they allow us to trade recognition accuracy and latency in a flexible manner.

\newpage
\bibliographystyle{IEEEbib}
\bibliography{main}

\begin{thebibliography}{10}

\bibitem{vaswani2017attention}
Ashish Vaswani, Noam Shazeer, Niki Parmar, Jakob Uszkoreit, Llion Jones,
  Aidan~N Gomez, {\L}ukasz Kaiser, and Illia Polosukhin,
\newblock ``Attention is all you need,''
\newblock in {\em Advances in neural information processing systems}, 2017, pp.
  5998--6008.

\bibitem{dai2019transformer}
Zihang Dai, Zhilin Yang, Yiming Yang, William~W Cohen, Jaime Carbonell, Quoc~V
  Le, and Ruslan Salakhutdinov,
\newblock ``Transformer-xl: Attentive language models beyond a fixed-length
  context,''
\newblock in {\em Proceedings of the 57th Annual Meeting of the Association for
  Computational Linguistics}, 2019, p. 2978–2988.

\bibitem{irie2019language}
Kazuki Irie, Albert Zeyer, Ralf Schlüter, and Hermann Ney,
\newblock ``{Language Modeling with Deep Transformers},''
\newblock in {\em Proc. Interspeech}, 2019, pp. 3905--3909.

\bibitem{transformerasr:01}
Linhao Dong, Shuang Xu, and Bo~Xu,
\newblock ``Speech-transformer: A no-recurrence sequence-to-sequence model for
  speech recognition,''
\newblock in {\em Proc. ICASSP}, 04 2018, pp. 5884--5888.

\bibitem{transformerasr:02}
Ngoc{-}Quan Pham, Thai{-}Son Nguyen, Jan Niehues, Markus M{\"{u}}ller, and Alex
  Waibel,
\newblock ``Very deep self-attention networks for end-to-end speech
  recognition,''
\newblock {\em CoRR}, vol. abs/1904.13377, 2019.

\bibitem{transformermt:01}
Qiang Wang, Bei Li, Tong Xiao, Jingbo Zhu, Changliang Li, Derek~F. Wong, and
  Lidia~S. Chao,
\newblock ``Learning deep transformer models for machine translation,''
\newblock {\em CoRR}, vol. abs/1906.01787, 2019.

\bibitem{transformerasr:03}
``Syllable-based sequence-to-sequence speech recognition with the transformer
  in mandarin chinese,''
\newblock in {\em Proc. Interspeech 2018}. pp. 791--795, ISCA.

\bibitem{transformerasr:04}
Abdelrahman Mohamed, Dmytro Okhonko, and Luke Zettlemoyer,
\newblock ``Transformers with convolutional context for {ASR},''
\newblock {\em CoRR}, vol. abs/1904.11660, 2019.

\bibitem{Povey2018ATS}
Daniel Povey, Hossein Hadian, Pegah Ghahremani, Ke~Li, and Sanjeev Khudanpur,
\newblock ``A time-restricted self-attention layer for asr,''
\newblock {\em 2018 IEEE International Conference on Acoustics, Speech and
  Signal Processing (ICASSP)}, pp. 5874--5878, 2018.

\bibitem{Dong_2019}
Linhao Dong, Feng Wang, and Bo~Xu,
\newblock ``Self-attention aligner: A latency-control end-to-end model for asr
  using self-attention network and chunk-hopping,''
\newblock {\em 2019 IEEE International Conference on Acoustics, Speech and
  Signal Processing (ICASSP)}, May 2019.

\bibitem{Sperber_2018}
Matthias Sperber, Jan Niehues, Graham Neubig, Sebastian Stüker, and Alex
  Waibel,
\newblock ``Self-attentional acoustic models,''
\newblock {\em Proc. Interspeech}, Sep 2018.

\bibitem{tsunoo2019online}
Emiru Tsunoo, Yosuke Kashiwagi, Toshiyuki Kumakura, and Shinji Watanabe,
\newblock ``Towards online end-to-end transformer automatic speech
  recognition,''
\newblock {\em arXiv:1910.11871}, 2019.

\bibitem{Sak:14a}
Ha{\c{s}}im Sak, Andrew Senior, and Francoise Beaufays,
\newblock ``{Long Short-Term Memory Recurrent Neural Network Architectures for
  Large Scale Acoustic Modeling},''
\newblock in {\em Proc. Interspeech}, 2014.

\bibitem{graves:12}
Alex Graves,
\newblock ``Sequence transduction with recurrent neural networks,''
\newblock in {\em Proceedings of the 29th International Conference on Machine
  Learning}, 2012.

\bibitem{rao2017exploring}
Kanishka Rao, Ha{\c{s}}im Sak, and Rohit Prabhavalkar,
\newblock ``Exploring architectures, data and units for streaming end-to-end
  speech recognition with rnn-transducer,''
\newblock in {\em 2017 IEEE Automatic Speech Recognition and Understanding
  Workshop (ASRU)}. IEEE, 2017, pp. 193--199.

\bibitem{he2017asru}
Yanzhang~(Ryan) He, Rohit Prabhavalkar, Kanishka Rao, Wei Li, Anton Bakhtin,
  and Ian McGraw,
\newblock ``Streaming small-footprint keyword spotting using
  sequence-to-sequence models,''
\newblock in {\em Automatic Speech Recognition and Understanding (ASRU), 2017
  IEEE Workshop on}, 2017.

\bibitem{sak:17}
Ha{\c{s}}im Sak, Matt Shannon, Kanishka Rao, and Françoise Beaufays,
\newblock ``Recurrent neural aligner: An encoder-decoder neural network model
  for sequence to sequence mapping,''
\newblock in {\em Proc. Interspeech}, 2017, pp. 1298--1302.

\bibitem{jaitly:15}
Navdeep Jaitly, David Sussillo, Quoc~V Le, Oriol Vinyals, Ilya Sutskever, and
  Samy Bengio,
\newblock ``A neural transducer,''
\newblock {\em arXiv preprint arXiv:1511.04868}, 2015.

\bibitem{graves:06}
Alex Graves, Santiago Fern{\'a}ndez, Faustino Gomez, and J{\"u}rgen
  Schmidhuber,
\newblock ``Connectionist temporal classification: {Labelling} unsegmented
  sequence data with recurrent neural networks,''
\newblock in {\em Proc. ICML}, 2006.

\bibitem{tripathi:19a}
Anshuman Tripathi, Han Lu, Hasim Sak, and Hagen Soltau,
\newblock ``{Monotonic Recurrent Neural Network Transducer and Decoding
  Strategies},''
\newblock in {\em Proc. ASRU}, 2019.

\bibitem{rabiner:93}
L.~R. Rabiner and B.-H. Juang,
\newblock {\em Fundamentals of Speech Recognition},
\newblock PTR Prentice-Hall, Inc., Englewood Cliffs, New Jersey 07632, 1993.

\bibitem{Wang2019hybrid}
Yongqiang Wang, Abdelrahman Mohamed, Duc Le, Chunxi Liu, Alex Xiao, Jay
  Mahadeokar, Hongzhao Huang, Andros Tjandra, Xiaohui Zhang, Frank Zhang,
  Christian Fuegen, Geoffrey Zweig, and Michael~L. Seltzer,
\newblock ``Transformer-based acoustic modeling for hybrid speech
  recognition,''
\newblock {\em arXiv:1910.09799}, 2019.

\bibitem{park2019specaugment}
Daniel~S Park, William Chan, Yu~Zhang, Chung-Cheng Chiu, Barret Zoph, Ekin~D
  Cubuk, and Quoc~V Le,
\newblock ``Specaugment: A simple data augmentation method for automatic speech
  recognition,''
\newblock {\em arXiv preprint arXiv:1904.08779}, 2019.

\bibitem{panayotov2015librispeech}
Vassil Panayotov, Guoguo Chen, Daniel Povey, and Sanjeev Khudanpur,
\newblock ``Librispeech: an asr corpus based on public domain audio books,''
\newblock in {\em Proc. ICASSP}. IEEE, 2015, pp. 5206--5210.

\bibitem{NIPS2011_4329}
Alex Graves,
\newblock ``Practical variational inference for neural networks,''
\newblock in {\em Advances in neural information processing systems}, 2011, pp.
  2348--2356.

\bibitem{chorowski:17}
Jan Chorowski and Navdeep Jaitly,
\newblock ``Towards better decoding and language model integration in sequence
  to sequence models,''
\newblock in {\em Proc. Interspeech}, 2017, pp. 523--527.

\end{thebibliography}

\end{document}